\documentclass{icrc2009}

\usepackage{graphicx}   
\usepackage[caption=false]{caption}    
\usepackage[font=footnotesize]{subfig} 
\usepackage{fixltx2e}

\usepackage{url}

\newcommand{\shorttitle}[1]%
{\markboth{Proceedings of the 31\MakeLowercase{$^{st}$} ICRC, {\L}\'{o}d\'{z} 2009}{#1} }
\newcommand{\etal}{\MakeLowercase{\textit{et al. }}} 

\def\apj{Ap. J.}

\def\aap{Astron. \& Astrophys.}

\def\nat{Nature}
\def\vFv{$\nu$F$_{\nu}$}

\begin{document}
\title{Fermi-LAT Observations of Gamma-Ray Bursts }

\author{\IEEEauthorblockN{Nicola Omodei\IEEEauthorrefmark{1}, on behalf of the Fermi LAT and GBM collaborations}
                            \\
\IEEEauthorblockA{\IEEEauthorrefmark{1}INFN, Sez. Pisa, Largo B.Pontecorvo, 3, 56100 Pisa, Italy}}

\shorttitle{Omodei,  \etal Fermi-LAT Observations of GRBs}
\maketitle

\begin{abstract}
 The Large Area Telescope (LAT) on the Fermi Gamma-ray Space Telescope observatory is a pair conversion telescope sensitive to gamma-rays over more than four energy decades, between 20 MeV and more than 300 GeV. 
Acting in synergy with the Gamma-ray Burst Monitor (GBM) - the other instrument onboard the mission - the LAT features unprecedented sensitivity for the study of gamma-ray bursts (GRBs) in terms of spectral coverage, effective area, and instrumental dead time.
We will review the main results from Fermi-LAT observation of GRB, presenting the main properties of GRBs at GeV energies. 
\end{abstract}

 
\section{Introduction}

High-energy emission from GRBs was first observed by the Energetic
Gamma-Ray Experiment Telescope (EGRET, covering the energy range from $30\;$MeV to $30\;$GeV) onboard the CGRO
satellite. Emission above 100~MeV was detected in few distinct cases. 
GRB~930131 showed high-energy emission that was consistent with an extrapolation
from the typical keV-MeV emission
\cite{som94}. Most important, evidence for an additional high
energy component up to 200~MeV with a different temporal behavior than
the keV-MeV component was discovered in GRB~941017 \cite{gon03}. High
energy emission for this GRB was observed for longer than 200 seconds
with a single spectral component (Band function~\cite{band93}) being ruled out to
explain the data.  Of great interest also is GRB~940217 for which
delayed high-energy emission was detected up to $\sim$$90$ minutes
after the BATSE GRB trigger, including an 18 GeV photon detected $\sim$$75$ minutes post-trigger
\cite{hur94}. Its temporal evolution is not correlated with the low-energy emission, and the wide band spectrum is inconsistent with a pure synchrotron model. 
More recently, the GRID instrument onboard Astro-rivelatore Gamma a Immagini LEggero (AGILE) detected 10 high-energy events with energies up to 300~MeV from GRB 080514B, in coincidence with its lower energy emission, with a significance of 3.0~$\sigma$ \cite{giu08}. Some of these events appear to arrive after the low energy emission.

Fermi can observe gamma-ray bursts, with an unprecedented effective area above 100 MeV. Thanks to the LAT, the number of gamma-ray bursts ever detected at high-energies has doubled in the first few months of the mission, providing more statistics and, for the first time, enough events at high-energy to perform a detailed temporal-spectral analysis. 
After ten months from its launch on June 11th 2008, the LAT instrument has firmly detected 8 GRBs at energies above 100~MeV: GRB 080825C (GCN 8183), GRB 080916C (GCN 8246), GRB 081024B (GCN 8407), GRB 081215 (GCN 8684), GRB 090217 (GCN 8903), GRB 090323 (GCN 9021), GRB 090328 (GCN 9077) and GRB 090510 (GCN 9334). A detailed analysis of GRB 080916C has been presented in~\cite{GRB080916C}, and also summarized by Tajima et al. in this proceedings.  This extremely bright GRB has shown that the $>$100 MeV emission is not only extended, but also is delayed with respect to the GBM emission. 
The spectral evolution of GRB 080916C during the prompt emission is always consistent with a Band function, for which the spectral parameters evolve with time, characterizing observed temporal behaviour. Moreover, a long lived emission in the LAT was found, lasting more than 20 minutes longer than seen with GBM.
In this contribution, we briefly report the observations of the other GRBs, with emphasis on the common observed behaviors.
\section{Large Area Telescope Gamma-Ray Bursts}
{\bf GRB 080825C}. The burst of the 25th August 2008, was the first GRB firmly detected by the LAT~\cite{LAT080825C-GCN}.
The GBM flight software triggered~\cite{GBM080825C-GCN} on the signal from GRB 080825C, localizing it at $\sim$$60^\circ$ from the LAT boresight at the time of the trigger, which puts it at the edge of the LAT field-of-view where the effective area drops by a factor of $\sim$~3.
\begin{figure*}[ht]
\centering
\includegraphics[width=5in]{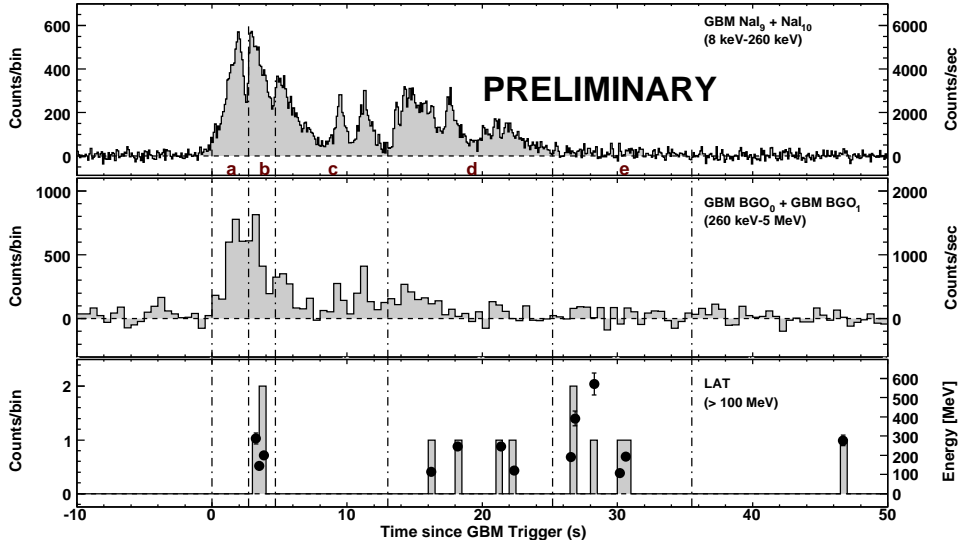}
\caption{Lightcurves of GRB 080825C observed by the GBM (NaI \& BGO) 
and LAT instruments; top two panels are background subtracted. The LAT 
lightcurve has been generated using selected events above 100 MeV (which are the events used for spectral analysis). Black dots, along with their error bars represent the $1$-$\sigma$ energy range for each LAT event.}
\label{080825C_LC}
\end{figure*}
\noindent
The top two panels of Fig. \ref{080825C_LC} show the background subtracted lightcurves of the two brightest NaI detectors (9 \& 10) and of the two BGO detectors. 
It exhibits a multiple peak structure with the two brightest peaks seen right after onset (interval ``a'' and ``b''). A series of less prominent peaks is well visible at interval ``c'' and ``d'' in both the NaI and BGO light curves. In interval ``e'' and after there is no longer emission recorded by the GBM.
The LAT ``transient'' selected events\footnote{Transient events are the events selected on the basis of quality cuts to match a background rate of $\sim$4 Hz above 100~MeV. See~\cite{atw08} for details.} above 100 MeV, detected close to the GBM position around the trigger time are shown in the bottom panel of Fig. \ref{080825C_LC}. All the events are compatible with the 95\% containment  radius, taking into account the energy dependence of the Point Spread Function~\cite{atw08}. The LAT data shows a count rate increase that is spatially and temporally correlated with the GBM emission (for a statistical significance of more than 6 sigma). 
From the LAT light curve, the emission above 100~MeV is apparently delayed by a few seconds (2.7 seconds, interval ``a'') but this delay is not statistically significant due to the lack of events. As measured in the GBM, the first peak is spectrally softer than subsequent peaks, and the LAT does not see any counts above 100 MeV during this time. The pulse in the GBM light curve at interval ``b'' is accompanied by LAT emission, while the three pulses during interval ``c'' are not. In Interval ``d'' both LAT and GBM emission are present, while interval ``e'', in which the LAT records the highest energy event for this burst (572$\pm$58~MeV), does not show evident GBM emission. 
A time resolved spectral analysis has been performed, combining GBM and LAT data. Fig.~\ref{080825C_LC} shows the resulting spectral model in the standard \vFv~representation. The figure shows the model that best fits the data in the each time bin, represented as a solid
line in the energy interval where data was used for the fit, and extrapolated up to 600~MeV, where the highest energy event was registered. The 68\% Confidence Level obtained using the covariance matrix is also shown.
In the first four intervals the spectral shape of GRB 080825C is consistent with a pure Band function, with the high-energy spectral index consistent with -2.5. In the last interval, where very small signal in the GBM is present, the spectrum is well fitted by a single power law, with a spectral index significantly harder than the values of the high-energy photon index $\beta$ of all of the earlier time bins.
The results presented here are still preliminary, and a detailed spectral analysis of this GRB will be presented in a forthcoming paper.
 

\begin{figure}[ht]
\centering
\includegraphics[width=3in]{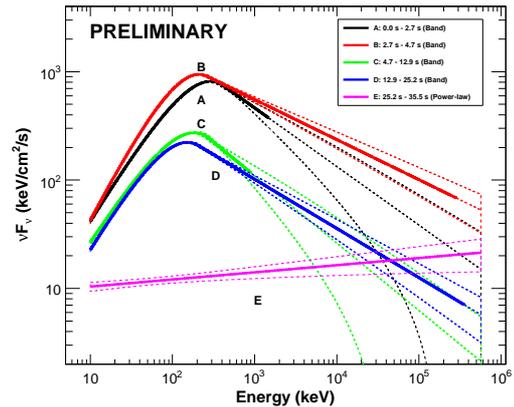}
\caption{Time-resolved spectral analysis results for GRB080825C. The various curves are the resulting models for the spectral analysis of the GRB 080825C. Continuous lines show the best fitted models in the energy bands used for the fits. They are extended up to the end of the LAT energy range with dashed lines. The dashed around the curves represent the Contour Level at the 68\% confidence interval. Each spectrum corresponds to the interval of the light curve of Fig.~\ref{080825C_LC}. For interval ``a,b,c,d'' the best fitting function adopted is a Band function, while in interval ``e'' a simple power law model provides better results.}
\label{080825C_SP}
\end{figure}
\noindent

\begin{figure*}[!t]
\centering
\includegraphics[width=5in]{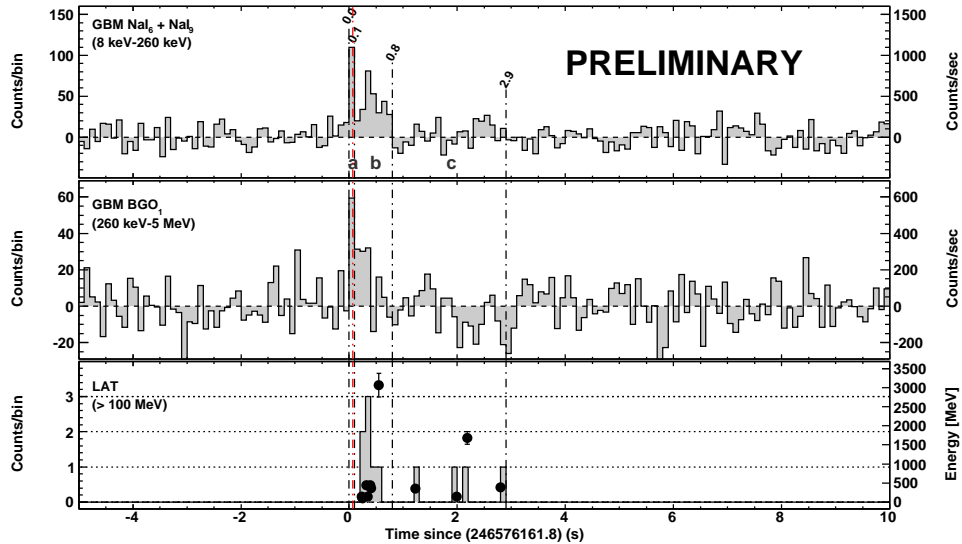}
\caption{Composite Light curves for the short burst 081024B. From top to bottom: the background subtracted light curve of the sum of two NaI detector; the background subtracted light curve of one BGO detector; the LAT events above 100~MeV. In this last panel, also the energy of the recorded events is shown, in MeV (right axis).}
\label{0801024B_LC}
\end{figure*}
\noindent
{\bf GRB 081024B}. The 24th of October 2008, the Fermi LAT telescope detected~\cite{LAT081024B-GCN} an increased count rate associated with the Fermi GBM GRB 081024B~\cite{GBM081024B-GCN}. The multi detector light curve is shown in Fig.\ref{0801024B_LC}. The top panel shows the background subtracted light curve for the sum of the signal of the two brightest NaI detectors. The time profile shows a narrow spike of about 0.1 seconds (interval ``a''), followed by a longer spike, of about 0.7 seconds (interval ``b''). There is no evidence of emission after $\sim$0.8 seconds from the trigger. 
The background subtracted light curve of the brightest BGO detector is shown in the second panel. Also here the time profile shows a narrow spike in coincidence with the NaI pulse, followed by a longer emission episode. In this case, due to the high background fluctuation and due to the fact that is not a very bright burst, it is hard to judge when the emission in the BGO ends, but, it is, apparently, compatible with, or even shorter, than the emission in the NaI. 
The third  panel of Fig.\ref{0801024B_LC} shows the LAT signal above 100~MeV, selecting events that belong to the ``transient'' event class selection, and compatible within the 95\% containment radius with the position of the GRB. 
Comparing the three panels we see that the first short pulse, clearly visible in the NaI and BGO light curves, is not present in the LAT data above 100~MeV, while the second broader emission period interval ``b'' is coincident with the first pulse in the LAT data. An event with energy $3\pm0.3$~GeV was detected 0.5 seconds after the trigger time while a second event at  1.7 GeV was detected 2.12 seconds post-trigger.
From the LAT light curve, it is evident that the emission above 100 MeV lasts up to approximately 3 seconds after the GBM trigger (interval ``c'') where 4 events are detected.
Though the statistics is too weak to make firm statements, the first pulse in the GBM light curve
is not present in the LAT, i.e. similarly to what has been observed for GRB 080825C and GRB 080916C (with very high significance in the latter case). 
This feature should not be seen as a time shift, since further peaks are ``in phase'', but as a signature of a typical spectral evolution.
We can use the background subtracted light curve to compute the duration of burst, by means of the standard parameter T$_{90}$ which expresses the time that the signal takes to go from  5\% to the 95\% of its intensity. 
The duration of the burst, computed using the integrated background subtracted NaI signal is about 0.7 seconds. This GRB triggered also the Suzaku Wide-band All-sky Monitor (WAM)~\cite{GCN8444} and the short-duration was confirmed also by this observation.
Using the signal from the LAT detector the duration of the burst is longer, with a T$_{90}$ of about 2.6 seconds. The short duration of the burst observed in the keV-MeV range, together with the extended LAT emission, makes this burst the first uncontroversial short-burst to have high-energy emission, with the extended nature of this emission providing new information on the nature of this class of bursts.
\begin{figure}[ht]
\centering
\includegraphics[width=3in]{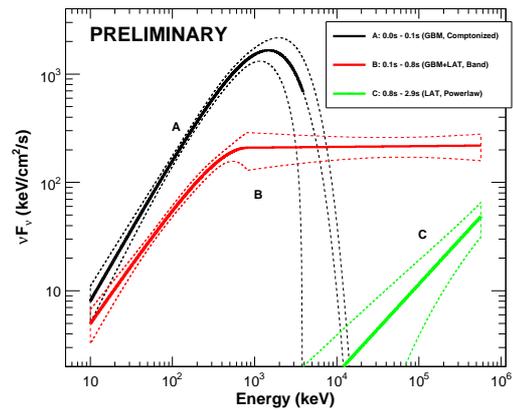}
\caption{Time-resolved spectral analysis results for GRB081024B.}
\label{081024B_SP}
\end{figure}
\noindent
A detailed time-resolved spectral analysis has been performed also for this burst. Fig.~\ref{081024B_SP} shows the three \vFv~spectra, each one corresponding to one of the intervals in the light curve. 
In the first time bin, the spectrum is well represented by a "Comptonized" model, which is a power law with an exponential cut-off. In this interval, also a Band function could be used to describe the data, but the high-energy spectral index would be very steep and poorly constrained. In the second time bin, the spectrum is well represented by a Band function with a high-energy spectral index close to -2. In the third time bin the GBM signal is very low and the spectrum is constrained with a simple power law (note that only four events contribute to this fit).

{\bf GRB 081215A} was detected by the GBM~\cite{GBM081215A-GCN}, which reported a very bright burst, with a T$_{90}$ of about 7.7 s in the 8-1000 keV. This burst was at an angle of 86 degrees to the LAT boresight, which  means that neither directional nor energy information can be obtained  with the standard analysis procedures. Using a non-standard data selection, over 100 counts above background were detected within a 0.5 s interval in coincidence with the main GBM peak.  
The significance of this excess in the LAT was greater than 8 sigma [14]. At very large angle, a
detection condition selects those events which scatter downwards and interact with at least three
planes of the LAT tracker, as required by the instrument trigger logic. This condition results in
the detection of low energy events only, for which the multiple scattering is large. A preliminary
study of the instrument performance at such a large inclination suggests that 95\% of the observed
events are likely to be gamma-rays with energies less than 140 MeV.

{\bf GRB 090217} was detected on ground by the LAT search algorithm, that automatically scans the LAT events looking for a significant excess associated with a clustering of events in the sky~\cite{LAT090217-GCN}. The LAT GRB was associated with the 32 seconds long GRB 090217 detected by the GBM~\cite{GBM090217-GCN}. The LAT emission above 100 MeV continues for up to 20 seconds after the GBM trigger, but commences several seconds after the GBM trigger.

{\bf GRB 090323} was detected by the GBM~\cite{LAT090323-GCN} that issued an Autonomous Repoint Request (ARR). The spacecraft immediately slewed to keep the GRB in the LAT field of view. This maneuver has been crucial for collecting data that we would have missed with the normal ``rocking'' profile. Emission from this burst was observed in the LAT up to a few GeV with a detection significance of more than 5 sigma. The high-energy emission commences several seconds after the GBM trigger time, and continues for up to a couple of kilo-seconds. 
A second, similar burst occurred few days later. 

The {\bf GRB 090328} was detected by the GBM, and also by the LAT ($>$5 $\sigma$) up to few GeV~\cite{LAT090328-GCN}. Also for this burst, the Fermi Observatory executed a maneuver following this trigger and tracked the burst location for the next 5 hours, subject to Earth-angle constraints. Further analysis with the LAT data~\cite{LAT090328-GCN2}, showed that the emission in the LAT lasts up around 900s after the trigger time.  
The long lasting high-energy emission detected by the LAT, with high-energy events positionally consistent with the GRB positions, makes these two cases extremely interesting.
Detailed analysis of these bursts will be presented in a forthcoming paper.

{\bf GRB 090510} is an extremely bright short burst, detected onboard by the LAT~\cite{LAT090510-GCN}, and by the GBM~\cite{GBM090510-GCN}. On ground analysis~\cite{LAT090510-GCN2} shows that more than 50 events above 100 MeV have been detected in the first second after the GBM trigger (more than 10 events have been detected above 1 GeV in the same interval). 
The LAT detector recorded also a long high-energy tail, lasting for about a minute, during which more than 150 events above 100 MeV ($>$20 above 1 GeV) were detected.
All these events are positionally consistent with the position of the GRB.

\section{Conclusions}
A new era for the gamma-ray astrophysics has begun. 
After a few months from its launch, the LAT telescope has already doubled the number of gamma-ray bursts detected above 100 MeV. There are similar trends in the bursts observed so far, and we believe that the understanding of these new properties is needed for the overall understanding of gamma-ray burst phenomena. 
After the detection of eight GRB by the LAT, we are starting to discern that the onset of 100 MeV detection is delayed with respect to the detection of lower-energy photons.
Typically the onset appears to be characterized by a different high-energy spectral index with respect to the remaining part of the burst.
As proposed in~\cite{GRB080916C}, this can occur if the pulses originate in two distinct physical regions, with different physical conditions. In the framework of internal shocks,  this naturally happens as the two emission are related to different sets of shells. This could be
explained as an additional hard component that arises after and lasts longer than the GBM emission. This component emerges distinctly in the late part of the LAT signal, where the GBM light curve has decreased below detectability. 
The existence of an extra component needs to be proved with brighter bursts where the evolution of the spectrum can be constrained with better accuracy. For example, the evidence of an extra component (like in the case of the EGRET burst 941017) will be unveiled only when a significant deviation from the Band spectrum is observed.
Temporally extended emission also appears common in LAT gamma-ray bursts, suggesting that the
high-energy emission mechanism lasts longer than previously believed.

\clearpage

\end{document}